\documentclass[a4,12pt,elsart]{article}
\usepackage{graphics}
\usepackage{epsfig}
\usepackage{epsf}

\begin{document}
\title{Production networks and failure avalanches}

\author{
 G{\'e}rard~Weisbuch$^{*}$ \\
Stefano Battiston$^{**}$ \\
$^{*}$Laboratoire de Physique Statistique\footnote{
Laboratoire associ{\'e} au CNRS (URA 1306), {\`a} l'ENS et
 aux Universit{\'e}s Paris 6 et Paris 7}
\\
     de l'Ecole Normale Sup{\'e}rieure, \\
    24 rue Lhomond, F-75231 Paris Cedex 5, France. \\
$^{**}$Centre Analyse et Math\'ematique Sociale, EHESS, Paris \\
  {\small     {\em email}:weisbuch@lps.ens.fr}\\
 }
\maketitle

\abstract{Although standard economics textbooks 
are seldom interested in production networks,
 modern economies
are more and more based upon suppliers/customers 
interactions. 
 One can consider entire sectors of the economy
as generalised supply chains.
We will take this view in the present paper
and study under which conditions local failures
to produce or simply to 
 deliver can result in avalanches of shortage and bankruptcies
across the network. We will show that a large class
of models exhibit scale free distributions
of production and wealth among firms and that metastable
regions of high production are highly localised.}

\section{Networks of firms}
\label{intro}

   Firms are not simply independent agents
competing for customers on markets.
 Their activity involves many interactions,
and some of them even involve some kind of cooperation.
Interactions among firms might include:
\begin{itemize}
\item information exchange\cite{Davis}\cite{Bat03a},\cite{Bat03b};
\item loans\cite{stig},\cite{cats};
\item common endeavours\cite{pow};
\item partial ownership\cite{Bat03c};
\item  and of course economic transactions allowing 
production\cite{bak} (the present paper).
\end{itemize}
  Economic activity can be seen as occurring on
an economic network (``le tissu \'economique''):
firms are represented by vertices and their interactions
by edges. The edges are most often asymmetric
(think for instance of providers/customers interactions).
The availability of empirical data has provoked
research on the structure of these networks:
many papers discuss their ``small world properties''\cite{wat}
and frequently report scale free distribution\cite{BA} of the connections
among firms. 

The long term interest of economic network
research is rather the dynamics creating or occurring 
on these nets: how are connections evolving, what are the fluxes
of information, decisions\cite{Bat03a},\cite{Bat03b}, economic transactions
 etc ...
 But dynamic studies lag behind 
 statistical approaches because of conceptual difficulties
and because time series of individual transactions are 
harder to obtain than time aggregated statistics.

 The recent cascade of bankruptcies that occurred in Eastern
Asia in 1997,  provoked
some research on the influence of the loans network structure
 on the propagation of ``bad debts'' and resulting avalanches
of bankruptcies (\cite{stig},\cite{cats})
. One of the most early papers on avalanche distribution
in economic networks is due to Bak {\it et al}
\cite{bak}. It concerns production networks:
edges represent suppliers/customers connections
among firms engaged in batch production activity.
The authors describe the distribution of production
avalanches triggered by random independent demand
events at the output boundary of the production network.

These papers (\cite{stig},\cite{cats} and\cite{bak}) are not based on any empirical description
of the network structure, but assume a very simple interaction
structure: star structure in the first case\cite{stig},\cite{cats}, periodic
lattice in Bak  {\it et al} paper\cite{bak}.
They neither take into account price dynamics.

 The present paper is along these lines:
we start from a very simple lattice structure
and we study the consequences of simple local processes
of orders/production (with or without failure)/delivery/
profit/investment on the global dynamics: evolution of 
global production and wealth in connection
to their distribution and local patterns. 
In the spirit of complex systems analysis,
our aim is not to present specific
 economic prediction, but primarily to concentrate on the generic
properties (dynamical regimes, transitions, scaling laws)
common to a large class of models of production networks.

A minimal model of a production network will first 
be introduced in section 2. Simulation results are presented in 
section 3. Section 4 is a discussion 
of the genericity of the obtained results:
reference is made to comparable soluble models.
We also summarise the results of several variants 
of the simplest model. The conclusion is a 
discussion of possible applications to 
geographical economics.

\section{ A simple model of a production network}

  We can schematise the suppliers/customers interactions
among firms by a production network, where firms are
located at the vertices and directed edges represent
the delivery of products from one firm to its customers
(see figure 1).

 Independent local failures to produce (or to deliver) by a firm 
might give rise to the propagation of shortage across the
production network.

\begin{figure}[htbp]
\centerline{\epsfxsize=120mm\epsfbox{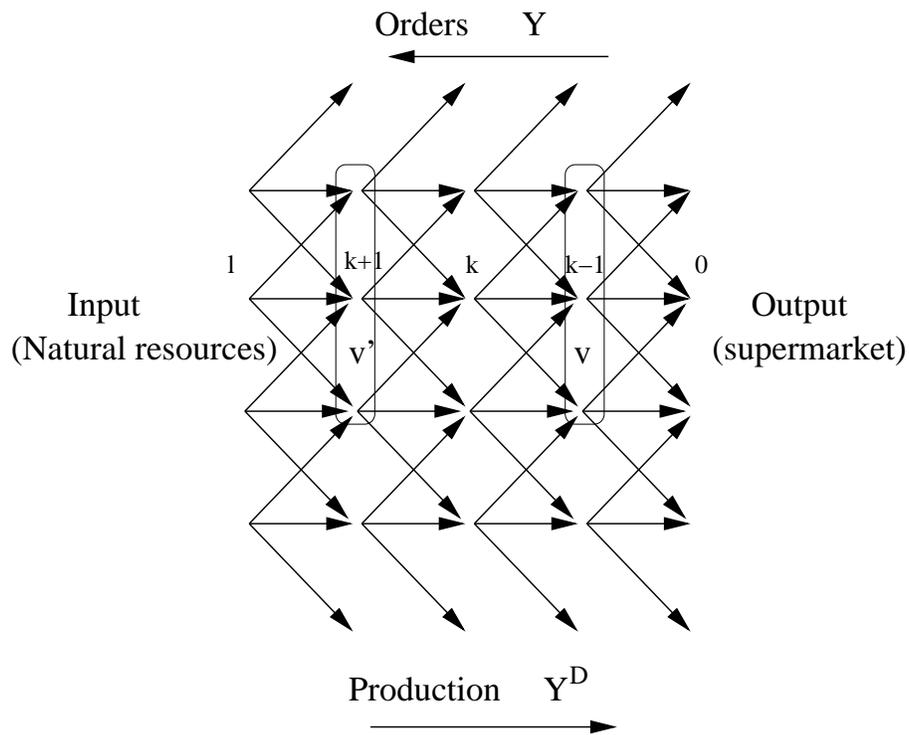}}
\caption{Firms are located at the nodes of the lattice.
Production ($Y^D$) flows from the resource input layer ($k=l)$ to
the output layer ($k=0$), orders ($Y$) flow backward.} 
\end{figure}

We have chosen a simple periodic lattice with three input connections
of equal importance and  three output per firm. The network is oriented
from an input layer (say natural resources) towards an output layer
(say the shelves of supermarkets). The transverse axis can be thought
as representing either geographical position or some 
product space while the longitudinal axis relates to production.
 We here use a one dimensional transverse space to facilitate
the representation of the dynamics by two-dimensional patterns,
but there is no reason to suppose
geographical or product space to be one-dimensional
 in the real world.   

In real economies, the network structure is
more heterogenous with firms of unequal importance and connectivity. 
Furthermore some delivery 
connections go backwards. Most often these backward 
connections concern equipment goods;
 neglecting them as we do here implies considering 
equipment goods dynamics as much slower
 than consumption goods dynamics.
Anyway, since these  backward 
connections enter positive feedback loops,
we have no reason to suppose that they would 
qualitatively disrupt the dynamics that we further describe.

At each time step two opposite flows get across the lattice: 
orders are first transmitted upstream from the output layer;
production is then transmitted downstream from the 
input layer to the output layer.

\begin{itemize}
\item Orders at the output layer

  We suppose that orders are only limited by the production 
capacity\footnote{A number of simplifying assumptions of our
model are inspired from \cite{cats}, especially the assumption that 
production is limited by production capacity, not by market.} 
$A_{0i}$ of the firm in position ${0,i}$, where $0$ indicates the output
layer, and $i$ the transverse position in the layer.
\begin{eqnarray}
  Y_{0i} &=& q \cdot A_{0i}
\end{eqnarray}
$Y_{0i}$ is the order in production units, and $q$ a 
technological proportionality coefficient relating the quantity 
of product $Y$ to the production capacity $A$, combining the
effect of capital and labor. $q$ is further taken equal to 1 without
loss of generality.

\item Orders 

Firms at each layer $k$, including the output layer, transfer orders
upstream to get products from layer $k+1$ allowing them to produce.
These orders are evenly distributed across their 3 suppliers upstream.
  But any firm can only produce according to its own production capacity 
$A_{ki}$. The planned production $Y_{ki}$ is then a minimum between
 production capacity and orders coming from downstream:
  \begin{eqnarray}
  Y_{ki} &=& min (q \cdot A_{ki} , \sum_{v}   \frac{Y_{(k-1)i}}{3})
\end{eqnarray}
 $v$ stands for the supplied neighborhood, here supposed to be the 
three firms served by firm $k,i$ (see figure 1).

  We suppose that resources at the input layer are always in excess
and here too, production is limited only by orders and production capacity.

\item Production downstream

 Starting from the input layer, each firm then starts producing
according to inputs and to its production capacity; but production
itself is random, depending upon alea. We suppose that at each time step
some catastrophic event might occur with constant probability $\mathcal{P}$ and
completely destroy production. Failures result in canceling production
at the firm where they occur, but also reduce production downstream, 
since firms downstream have to reduce their own production by lack of input.
These failures to produce are uncorrelated in time and location on
the grid.   
 Delivered production  $Y^d_{ki}$ by firm $k,i$ then depends upon the
production delivered upstream from its delivering neighborhood $v'_i$
at level $k+1$:
\begin{eqnarray}
  Y^d_{ki} &=&  (\sum_{i'\in v'_i} Y^d_{(k+1)i'} \cdot \frac{Y_{ki}}{\sum_{i''\in v_{i'}} Y_{ki''}} ) \cdot \epsilon(t) 
\end{eqnarray}
\begin{itemize}
\item  Whenever any of the firms $i'\in v'_i$ at level $k+1$ is not able to deliver 
according to the order
it received, it delivers downstream at level $k$ 
to its delivery neighbourhood $v_{i'}$
in proportion of the initial orders
it received, which corresponds to the fraction term; 
\item 
 $\epsilon(t)$ is a random term equals to 0 with probability $\mathcal{P}$
and 1 with probability $1-\mathcal{P}$.
\end{itemize}
  The propagation of production deficit due to local 
independent catastrophic event is the collective
phenomenon we are interested in.

\item Profits and production capacity increase

  Production delivery results into payments without failure.
For each firm, profits are the difference between the valued quantity
of delivered products and production costs, minus capital decay.
 Profits $\Pi_{ki}$ are then written:
\begin{eqnarray}
  \Pi_{ki} &=& p\cdot Y^d_{ki} - c \cdot Y^d_{ki} - \lambda A_{ki}
\end{eqnarray}
where $p$ is the unit sale price,
 $c$ is the unit cost of production,
and $\lambda$ is the capital decay constant due to interest rates and
material degradation.
We suppose that all profits are re-invested into production.
Production capacities of all firms are
 thus upgraded (or downgraded in case
of negative profits) according to:
\begin{eqnarray}
A_{ki}(t+1)=A_{ki}(t)+ \Pi_{ki} (t)
\end{eqnarray}

\item Bankruptcy and re-birth.

  We suppose that firms which capital becomes negative go
into bankruptcy. Their production capacity goes to zero
and they neither produce nor deliver. In fact
 we even destroy firms which
capital is under a minimum fraction of the average firm (typically 1/500).
  A re-birth process occurs for the corresponding vertex after 
a latency period: re-birth is due to the creation of new firms which use
the business opportunity to produce for the downstream neighborhood
of the previously bankrupted firm. New firms are created at a
unique capital, a small fraction of the average firm capital (typically
1/250).\footnote{Adjusting these capital values relative to the average
firm capital  $<A>$ is a standard hypothesis in many economic growth 
models: one supposes that in evolving economies such processes depend upon
the actual state of the economy\cite{solGLV} and not upon fixed and predefined values.}.
\end{itemize}

  The dynamical system that we defined here belongs to a large
class of non linear systems called reaction-diffusion systems (see e.g. \cite{solAB})
from chemical physics. The reaction part here is the autocatalytic loop
of production and capital growth coupled with capital decay and death 
processes. The diffusion part is the diffusion of orders and production 
across the lattice. We can a priori expect a dynamical behaviour 
with spatio-temporal patterns, well characterised 
dynamical regimes separated in the 
parameter space by transitions or crossovers, and scale free distributions
since the dynamics is essentially multiplicative and noisy.
These expectations guided our choices of quantities to monitor
during simulations.

\section{ Simulation results}

\subsection{ Methods and parameter choice}

   Unless otherwise stated, the following results were obtained for a 
production network with 1200 nodes and ten layers between the input and
the output. 

Initial wealth is uniformly and randomly distributed 
among firms:
\begin{equation}
  A_{ki} \in [1.0,1.1]
\end{equation}

One time step correspond to the double sweep
of orders and production across the network,
plus updating capital according to profits.
The simulations were run for typically 5000 time steps.

The figures further displayed correspond to:
\begin{itemize}
\item  a capital threshold for bankruptcy of $<A>/500$;
\item  an initial capital level of new firms of $<A>/250$;
\end{itemize}

Production costs $c$ were 0.8 and capital decay
rate $\lambda=0.2$. In the absence of failures,
stability of the economy would be ensured by
sales prices $p=1.0$. In fact, only the relative difference between 
these parameters influences stability. But their relative magnitude
 with respect to the inverse delay between 
bankruptcy and creation of new firm also
qualitatively influence the dynamics. 

In the limits of low probability of failures,
when bankruptcies are absent,
the linear relation between failure probability $\mathcal{P}$ 
and equilibrium price $p$ is
written:
\begin{eqnarray}
  p=c+\lambda+ \frac{l}{2} \cdot \mathcal{P} 
\end{eqnarray}
where $l$ is the total number of layers.
 The $\frac{l}{2}$ comes from the fact that the integrated damage
due to an isolated failure is proportional to the average number of
downstream layers. The slopes at the origin of
 the breakeven lines of figure 2 verify this equation.

  Most simulations were monitored online:
we directly observed the evolution of the
 local patterns of wealth and production
which our choice of a lattice topology made possible.
Most of our understanding comes from these direct observations.
But we can only display global dynamics or static patterns
 in this manuscript.

\subsection{Monitoring global economic performance}

The performance of the economic system under 
failures can be tested by checking which prices
correspond to breakeven: the capital dynamics being essentially
exponential, the parameter space is divided in two regions,
where economic growth or collapse are observed. 
Drawing the breakeven manifolds 
for instance in the failure probability $\mathcal{P}$ and 
sale price $p$ plane
allows to compare the influence of other parameters 
. The growth regime is observed in the low $\mathcal{P}$
and high $p$ region, the collapse regime in the 
high $\mathcal{P}$ and low $p$ region.

\begin{figure}[htbp]
\centerline{\epsfxsize=120mm\epsfbox{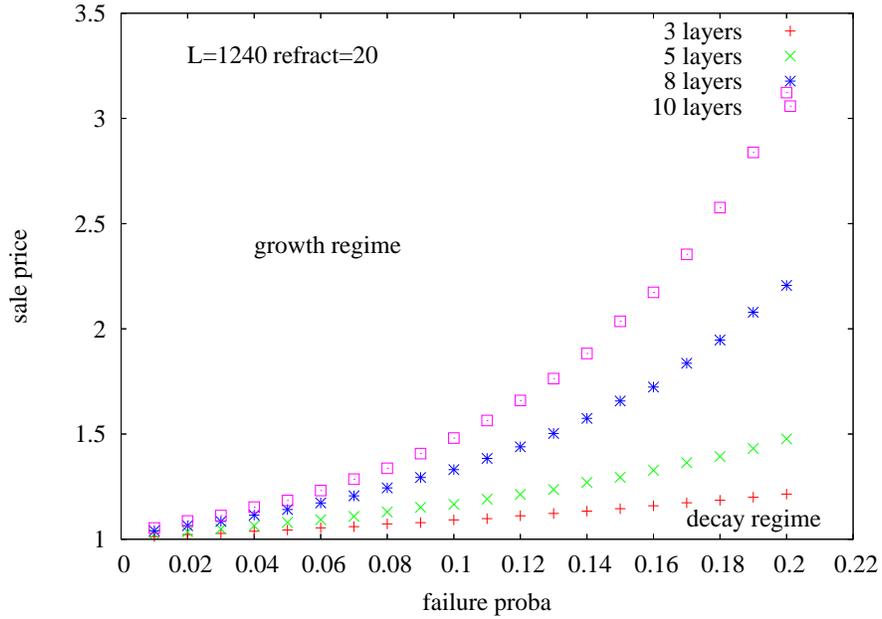}}
\caption{Regime diagram in the sale price versus
probability of failure plane. The time lag between bakruptcy
and re-birth is 20.
Two regions of growth and 
economical collapse at large times are separated by lines
which position are fixed by simulation parameters.
We here varied the production network depth:
 The red '+' line was obtained for a 3 layers net,
the green 'x' line for a 5 layers net,
 the blue '*' line for an 8 layers net,
and pink square line for a 10 layers net.
} 
\end{figure}

  Figure 2 displays four
breakeven manifolds corresponding to different lattice
depths.

At low failure probability, the breakeven lines follow equation 7.
At higher values of $\mathcal{P}$, interactions among
firms failures are important, hence the 
non linear increase of compensating prices.

 Breakeven manifold are a simple test of the economic 
performances of the network: when performances are poor, the compensating
sales price has to be larger. We checked for instance that increasing
the bankruptcy threshold and new firms initial capital increase
global economic performance. On the other hand, increasing 
the time lag between bankruptcy and the apparition of new firms
increase breakeven sale prices in the non-linear region.

Among other systematic tests, we checked parent models 
with more realistic representations of production costs such as:
\begin{itemize}
\item 
Influence of capital inertia; production costs don't instantly 
readjust to orders: capital and labour have some inertia
which we modeled by writing that productions costs are a maximum
function of actual costs and costs at the previous period.
\item  
Influence of the cost of credit: production failures increase
credit rates.
\end{itemize}
 Both variants of course yield higher breakeven sale prices;
nevertheless these variants display the same generic properties
that we will discuss in the next sections. 

 Most further results, dynamical and statistical,
are based on runs close to the breakeven price
in order to avoid systematic drifts and recalibrations.

\subsection{Time evolution}

 The simplest way to monitor the evolution of the system
is to display the time variations of some of its 
global performance. Figure 3 displays the time variations
of total delivered production $Y^d$, total wealth $A$,
total undelivered production due to failures
and the fraction of active firms for a 1200x10 lattice,
with a probability of failure of 0.05 and a compensation
sale price of 1.185. Time lag between bankruptcy and 
and new firm creation is either 1 (for the left diagram)
or 5 (for the right digram).

\begin{figure}[htbp]
\centerline{\epsfxsize=100mm\epsfbox{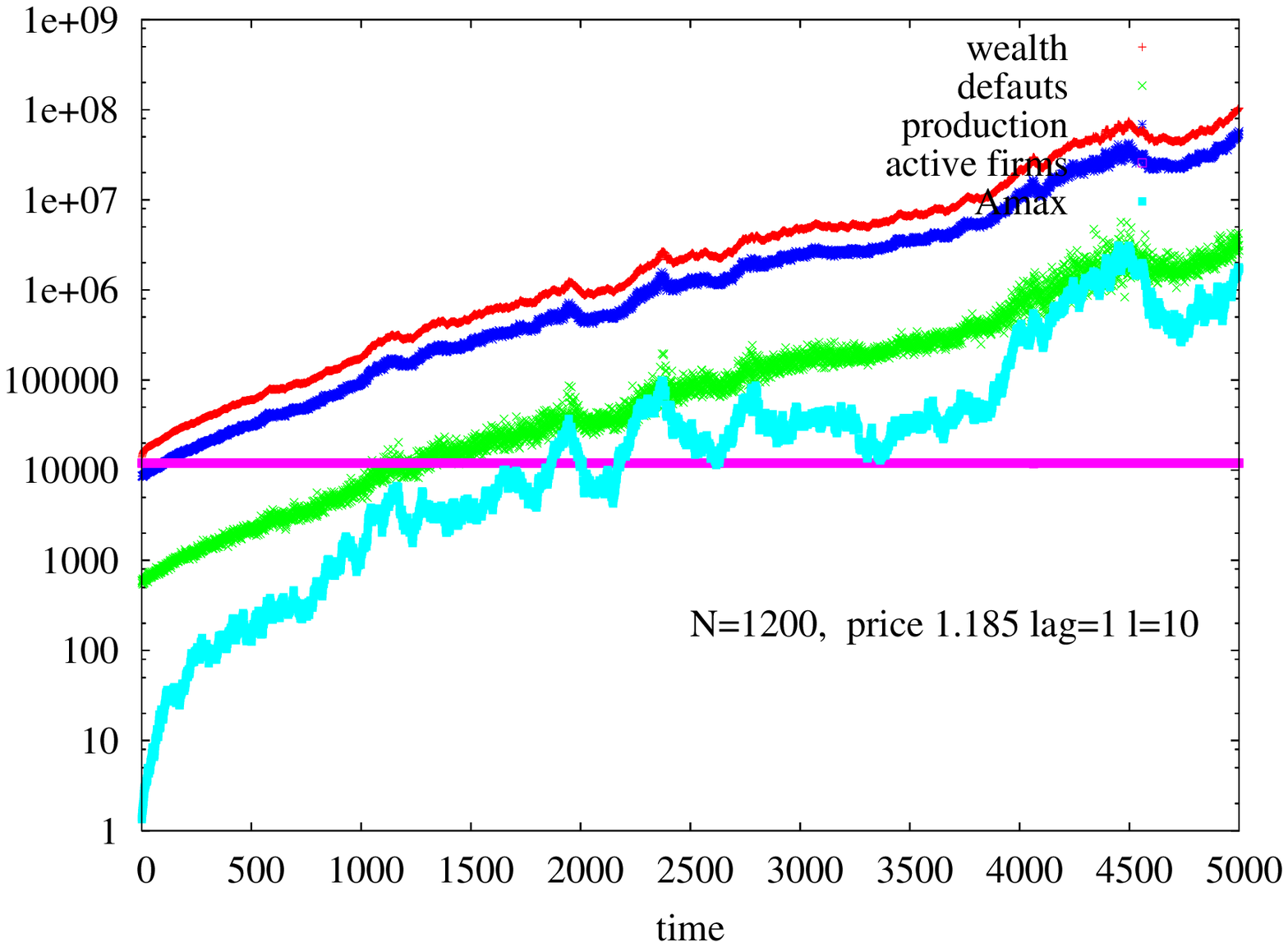}\epsfxsize=100mm\epsfbox{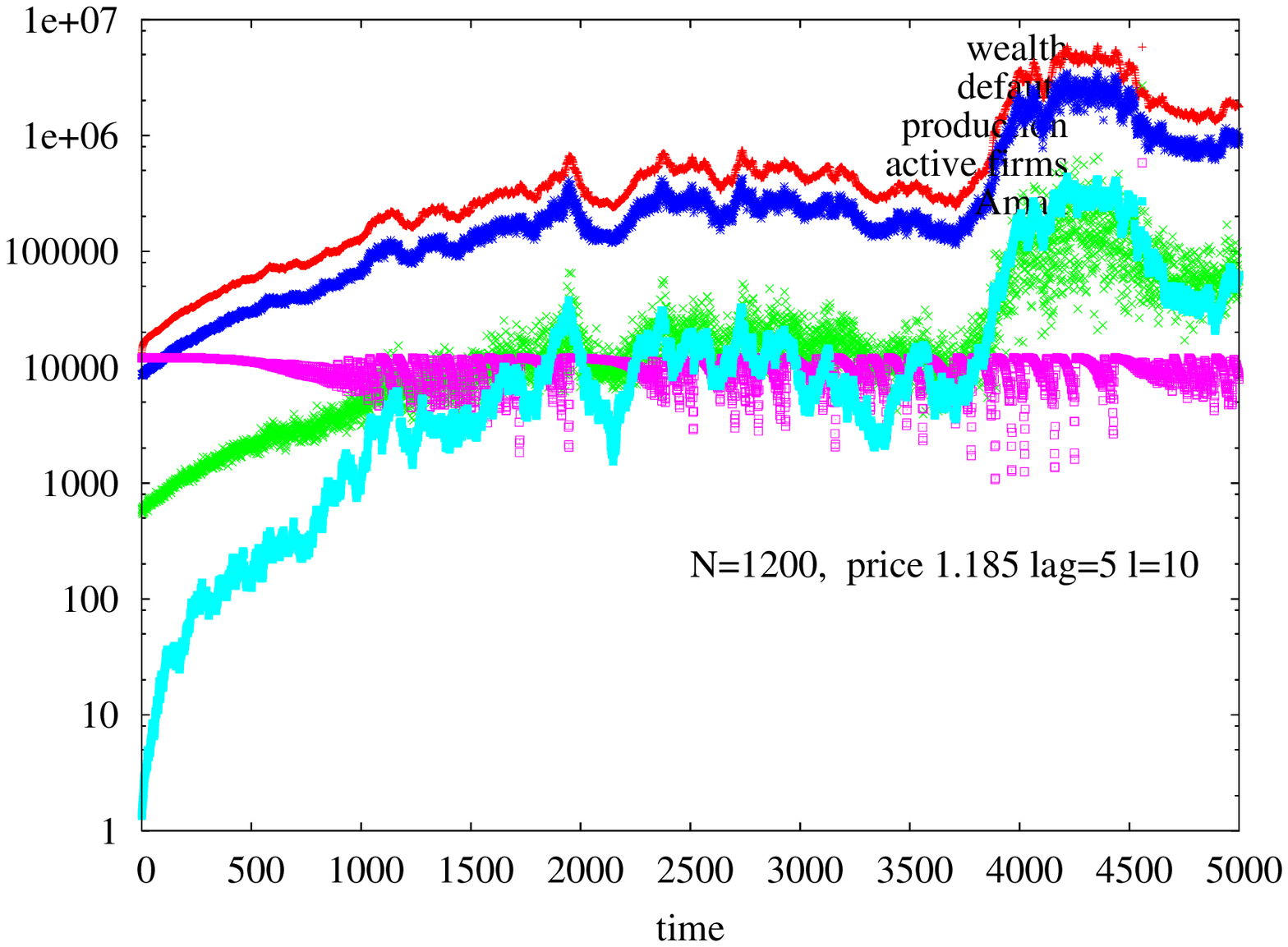}}
\caption{Time evolution of wealth (red '+'), production (blue '*'),
destroyed production (green 'x'), active firms (magenta empty squares) and
production by the largest firm (cyan hollow squares).
The network has 10 layers, 200 firms per layer, $\mathcal{P}=0.05$ (the
failure probability).
The left diagram corresponds to a small time lag (1) 
between bankruptcy and firm re-birth, right diagram corresponds to a
larger time lag (5). 
Vertical scale is logarithmic, which permits to have the four quantities
displayed on the same time plot but reduces the apparent amplitude of fluctuations 
occurring when time is larger than 1000.} 
\end{figure}

The features that we here report are generic to most
simulation at breakeven prices.
During the initial steps of the simulation, here say 1000, the wealth distribution
widens due to the influences of failures. Bankruptcies cascades
 do not occur as observed by checking the number of active firms,
until the lowest wealth values reach the bankruptcy threshold.
All quantities have smooth variations.
Later, for $t>1000$ one observes large production
 and wealth fluctuations characteristic
of critical systems.

At larger time lag (5) between bankruptcy and firm re-birth,
 when bankruptcies become frequent, they can cascade
across the lattice and propagate in both network directions
as seen on the right diagram of figure 3.
A surprising feature of the dynamics
is that avalanches of bankruptcies are not correlated with production level.
Even when only one tenth of the firms are active, the
total production is still high. In fact, in this model,
most of the total production
is dominated by large firms, and avalanches which concern
mostly small firms are of little consequence
for the global economy.

  Battiston etal study more thoroughly the time dynamics
of a related model (large sale price fluctuations
possibly inducing bankruptcies and lack of 
payment) in \cite{bat-galleg} .

\subsection{Wealth and production patterns}

Like most reaction-diffusion systems, the dynamics is
not uniform in space and display patterns.
The wealth and production patterns displayed after 5000 time steps
on figure 4 and 5 were obtained for  $\mathcal{P}=0.05$ .
They reflect wide 
 distributions and spatial organisation. 
In these diagrams, production flows upward. The upper diagram displays
wealth $A$ and the lower one production $Y_d$. The intermediate bar is
the colour scale, black=0, violet is the maximum wealth or production.
(We in fact displayed square roots of $A$ and $Y_d$ in order to 
 increase the visual dynamics of the displays; otherwise 
large regions of the patterns
would have been red because of the scale free distributions of $A$ and $Y_d$,
see further).

The important result is that although production has random fluctuations
and diffuses across the lattice, the inherent multiplicative
(or autocatalytic)  
process of production + re-investment coupled with 
local diffusion results in a strong metastable  
local organisation: the dynamics clusters rich and productive
firms in "active regions" separated by "poor regions" (in red or black).

\begin{figure}[htbp]
\centerline{\includegraphics[width=18cm, clip=true, trim= 0 0 0 3]{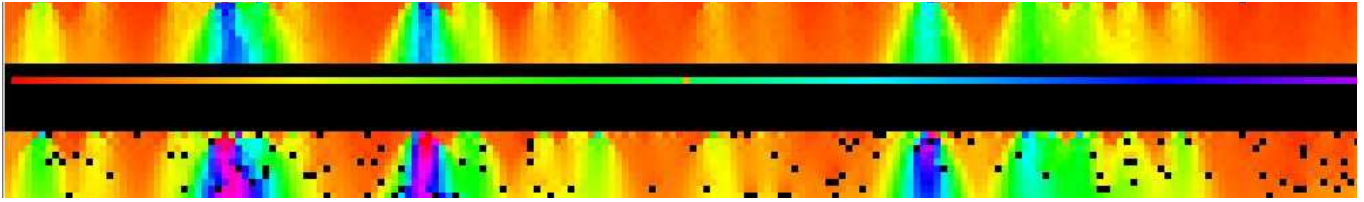}}
\caption{Patterns of wealth(upper pattern) and 
production (lower pattern)
after 5000 iterations steps with the parameter set-up of figure 3 (left)
(time lag =1), for a 200x10 lattice.
. For both patterns the output layer is
the last one above. The intermediate line is the colour code, with minimal
amplitude at the extreme left. We observe alternance of highly productive
regions
(in pink, blue and green colour),
with less active regions (in red). Local production failures
represented by black dots are randomly distributed across the production pattern.
Only one bankrupted firm is observed on the wealth pattern.} 
\end{figure}

\begin{figure}[htbp]
\centerline{\includegraphics[width=18cm, clip=true, trim= 0 0 0 3]{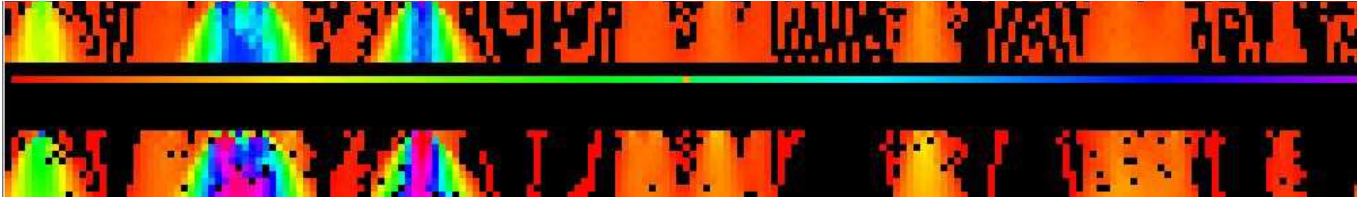}}
\caption{Patterns of wealth(upper pattern) and 
production (lower pattern)
after 5000 iterations steps with the parameter set-up of figure 3 (right)
(time lag is 5).
 The same alternance of active and less active regions is observed,
but with a larger time lag (5), we also get large zones of bankrupted firms
in black.} 
\end{figure}

\begin{figure}[htbp]
\centerline{\includegraphics[width=18cm, clip=true, trim= 0 0 0  3]{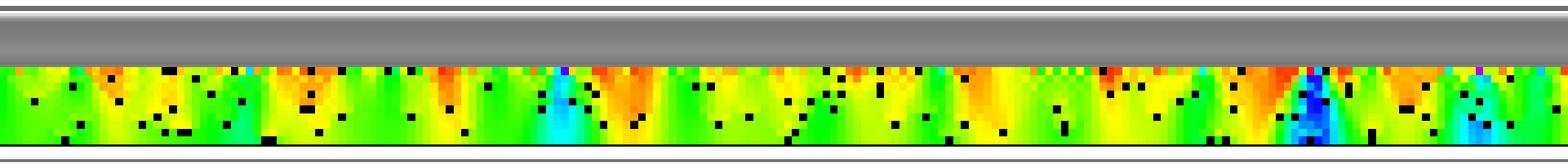}}
\centerline{\includegraphics[width=18cm, clip=true, trim= 0 0 0 3]{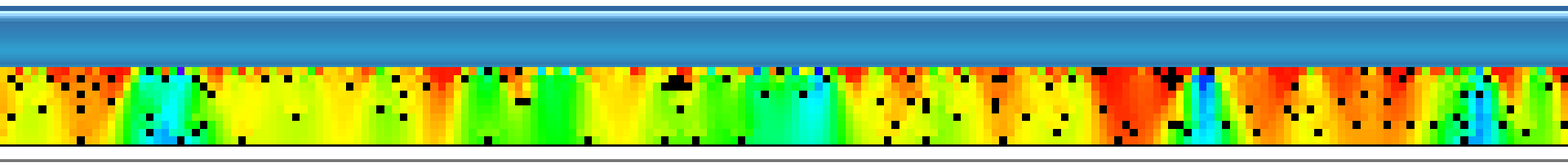}}
\centerline{\includegraphics[width=18cm, clip=true, trim= 0 0 0 3]{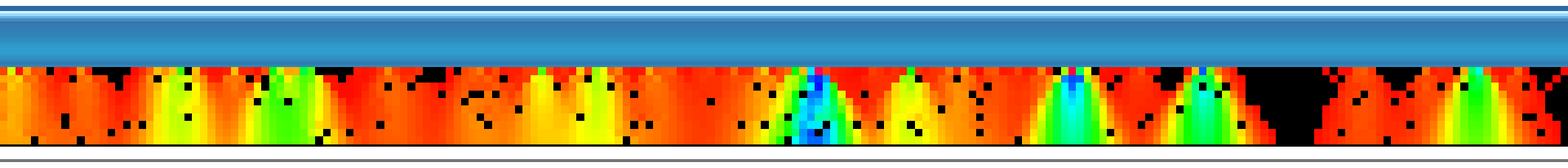}}
\centerline{\includegraphics[width=18cm, clip=true, trim= 0 0 0 3]{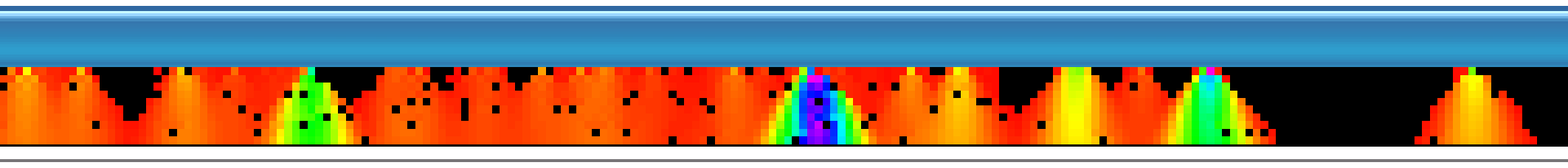}}
\centerline{\includegraphics[width=18cm, clip=true, trim= 0 0 0 3]{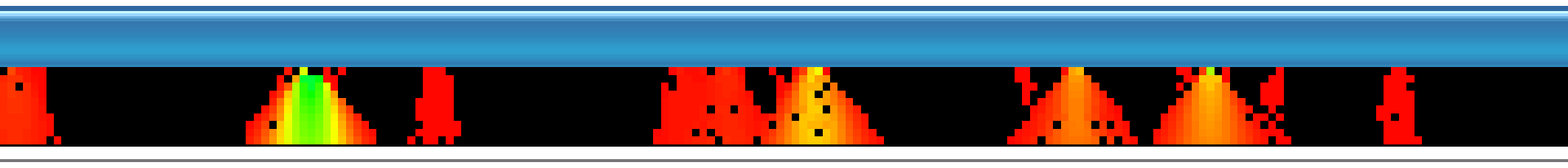}}
\caption{Successive patterns of wealth after 250, 750, 1250, 1750 and 2250
time steps with the parameter set-up of figure 3 (right, time lag = 5) for a 1200x10 lattice.}
\end{figure}

These patterns are evolving in time, but are
metastable on a long time scale, say of 
the order of several 100 time steps as seen on the succession of
production patterns at different steps of the simulation
as one can observe on figure 6: 
successive patterns at time 1250, 1750 and 2250.

  The relative importance of active (and richer)
regions can be checked by a Zipf plot\cite{Zipf}.
We first isolate active regions by "clipping" the dowstream
(along $k$ axis) integrated
wealth at a level of one thousandth of the total production\footnote
{Clipping here means that when the production level is lower than the threshold
it is set to zero}. 

\begin{figure}[htbp]
\centerline{\epsfxsize=120mm\epsfbox{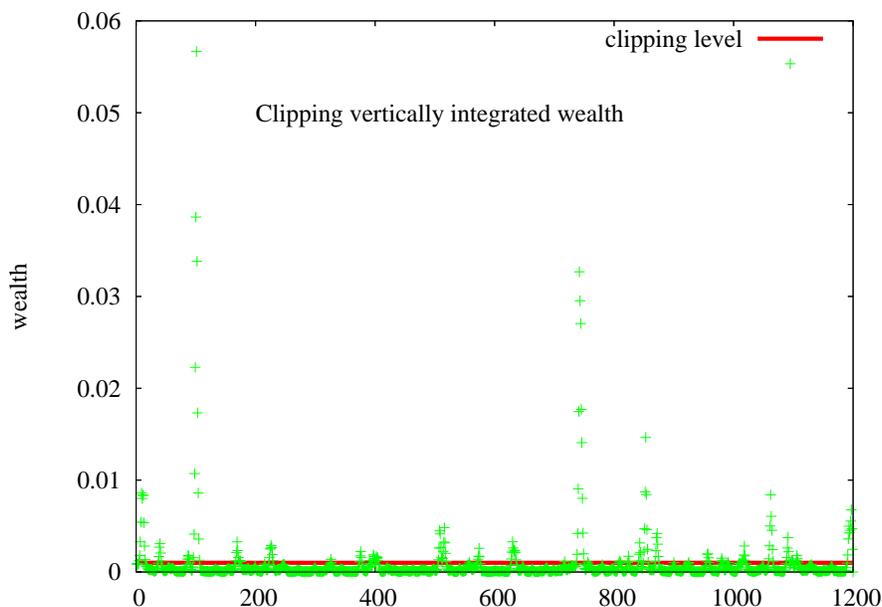}}
\caption{ Separating richer regions. Downstream integrated
wealth levels (green '+') are plotted as a function of their transverse position.
The clipping level indicated by the red line isolates richer regions
(those wealth peaks above the red line).}
 \end{figure}

We then transversally (along $i$ axis) integrate the wealth of active regions
and order these regional wealths to get the Zipf plots. 
\begin{figure}[htbp]
\centerline{\epsfxsize=150mm\epsfbox{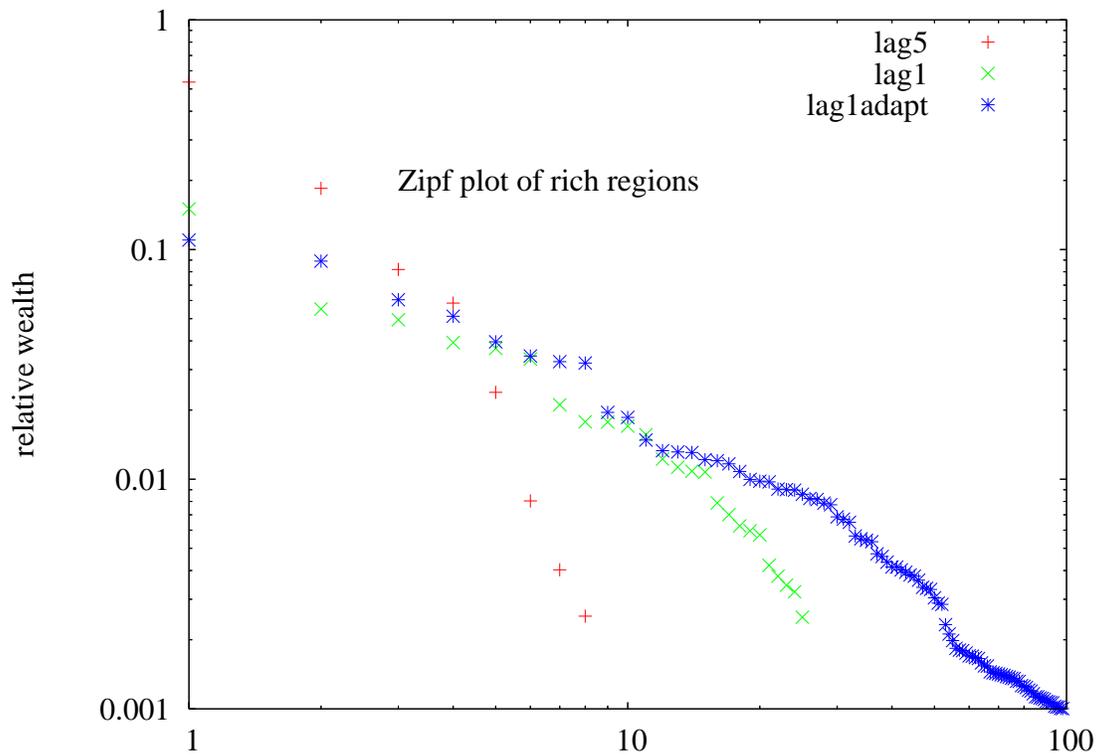}}
\caption{Zipf plot of wealth of the most active regions for 
the standard and adaptive firms models (cf. section 4.2).
The vertical axis display the production relative to 
the total production.
The red '+' correspond to the standard model with time lag = 5,
green 'x' to time lag = 1, and  blue '*' to the
 adaptive firms model with  time lag = 1.}
 \end{figure}

  All 3 Zipf plots display some resemblance with
standard Zipf\cite{Zipf} plots of individual wealth, firm size
and city size. For the model discussed
here, the size decrease following approximately a power law.
The apparent\footnote
{the approximate algorithm that we use to isolate
high productivity regions is responsible for the
kinks in the Zipf plot} exponent is one when the time lag is
1. It is much higher when  the time lag is
5. 

Zipf plots of output\footnote
{ rather than vertically integrating production,
we applyed the clipping, horizontal integration and
ordering algorithm to firms at the output layer ($k=0$)}
 active regions (not shown here) display
the same characteristics. 

When the time lag is 5, the most productive region
accounts for more than 50 perc. of total production.
The figure is 18 perc. for the second peak.
The distribution typically is "winner takes all".
The equivalent figures when the time lag is 1
are 10 and 8.5 perc..

 In conclusion, the patterns clearly display 
some intermediate scale organisation
in active and less active zones:
strongly correlated active regions are responsible 
for most part of the production. The relative
importance of these regions obeys a Zipf distribution.

\subsection{Wealth and production histograms}

  The multiplicative random dynamics of capital
and the direct observation of wealth and production
would lead us to predict a scale free distribution\footnote
{What we mean here by scale free is that no characteristic
scale is readily apparent from the distribution
as opposed for instance to gaussian distributions.
Power law distributions are scale free.
A first discussion of power law distributions
generated by multiplicative processes appeared in
\cite{kes}.}
 of 
wealth and production. 

 The cumulative distribution functions (cdf) of wealth and 
production  observed on figure 8 are indeed wide
range and do not display any characteristic scale:
The data wealth and production
were taken for the same conditions as the previous figures at the end of the
simulation, i.e. after 5000 time steps.
The medium range of the cdf
when time lag is 1 (figure 8a)
extends on one and a half decade with an apparent
slope of $1 \pm 0.05$ in log-log scale. 

This observed dependence of the wealth cdf, log normal at lower  $A$ values followed by
power law at intermediate $A$ values, is consistent with expressions derived 
for pdf in the literature on coupled differential equations with
multiplicative noise.
Bouchaud and M\'ezard\cite{bouch} e.g. obtained:
\begin{equation}
  P(w)= Z \frac{exp-\frac{1-\mu}{w}}{w^{1+\mu}} 
\end{equation}
(where $w$ stands for the wealth relative to average wealth $\bar{A}$),
from the differential system:
\begin{equation}
  \frac{dA_i}{dt}= \eta_i(t)\cdot A_i + J \cdot (\bar{A}-A_i) .
\end{equation}
where $\eta_i(t)$ is a random multiplicative noise,
with variance $\sigma^2$; $\mu=1+\frac{J}{\sigma^2}$.

At higher wealth, the straight line giggles
and drops much faster: this is because of the
underlying region structure. The last 80 perc.
of the wealth is concentrated in two rich regions
and its distribution is dominated by local
diffusion phenomena in these regions.

The departure form the standard (equ.8)
distribution is even more noticeable
when avalanches are present.
The large wealth shoulder is bigger
(95 perc. of production) and
the first point at zero wealth stands well above
 the rest of the distribution:
it corresponds to those 50 perc. of the firms
which are momentarily bankrupted.
 The fraction of 
bankrupted firms fluctuates
in time and so does the slope of the linear segment\footnote{
both fluctuations are correlated since the
slope of the linear segment depends upon the number of firms
in the distribution}.

\begin{figure}[htbp]
\centerline{\epsfxsize=100mm\epsfbox{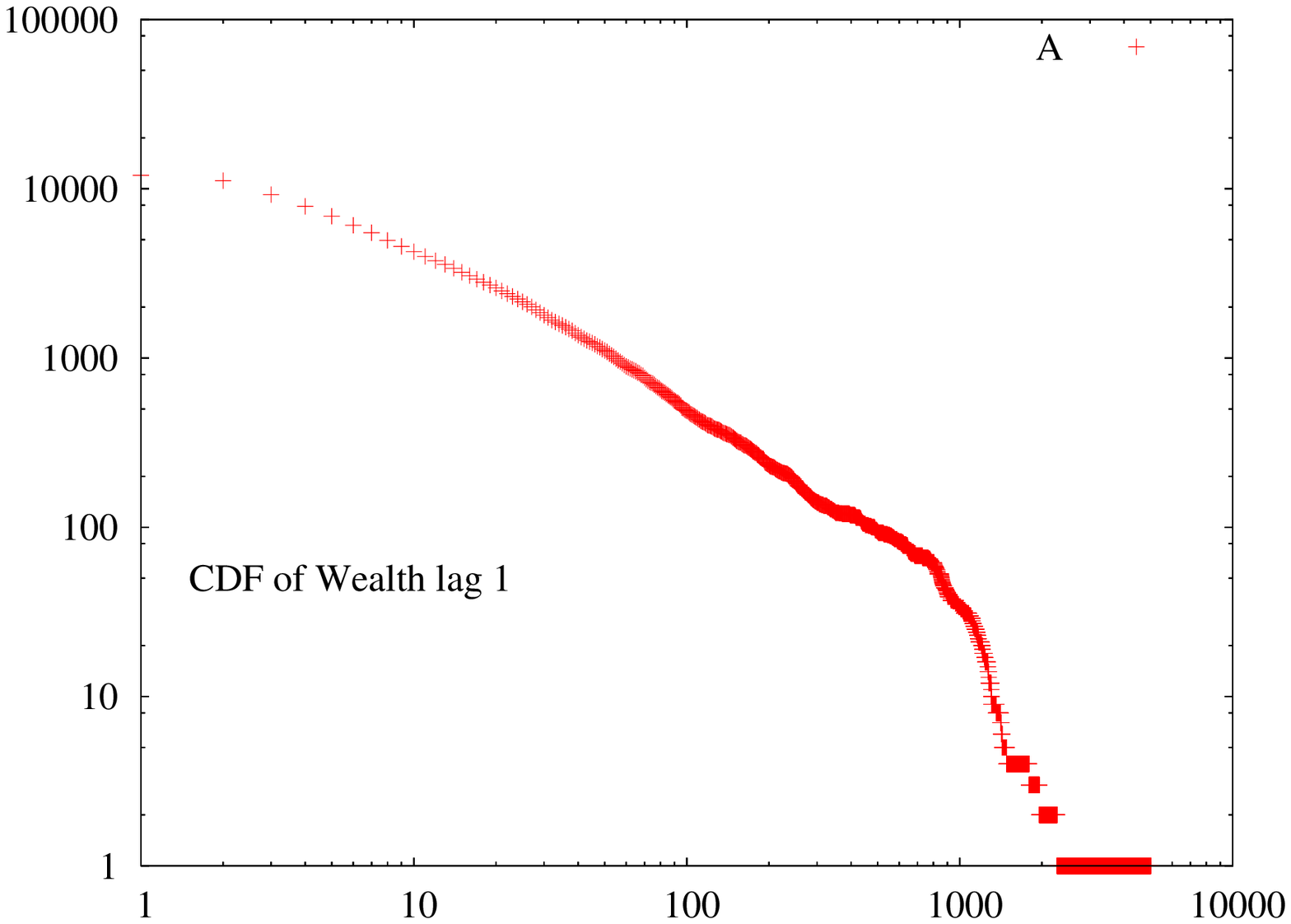}\epsfxsize=100mm\epsfbox{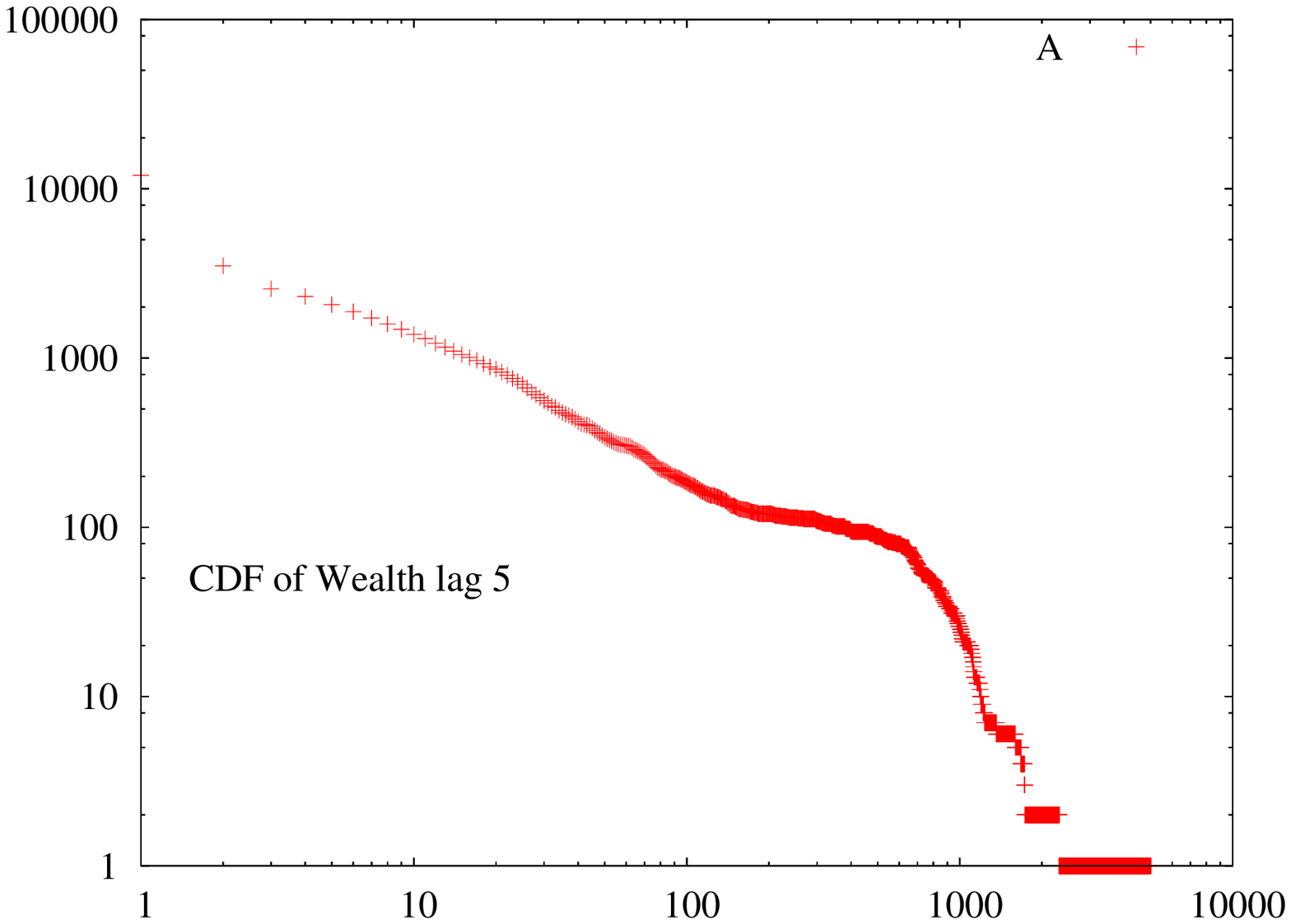}}
\caption{Cumulative distribution of wealth (red '+') after 5000 iteration steps. Parameter
choices are the same as the previous figures.} 
\end{figure}

In conclusion, the observed statistics largely reflect
the underlying region structure: at intermediate
levels of wealth, the different wealth peaks
overlap (in wealth, not in space!): we then observed
a smooth cdf obeying equation 8. At the 
large wealth extreme the fine structure of peaks is
revealed.

\section{Conclusions}

  The simple model of production networks that we 
proposed presents some remarkable properties:

\begin{itemize}
\item Scale free distributions of wealth and production.
\item Large spatial distribution of wealth and production.
\item A few active regions are responsible for most production.  
\item Avalanches of bankruptcies occur for larger values
of the time lag between  bankruptcy and firm re-birth.
But even when most firms are bankrupted, the global
economy is little perturbed.
\end{itemize}

Are these properties generic to a large class
of models? we will first briefly report on equations
which display similar behaviour and then examine
the results which we obtained with variants of the
model.

\subsection{Formal approaches of similar dynamics}

A number of models which display equivalent 
phenomena have been proposed and formally solved.
We kept our own notation to display similarities:

\begin{itemize}
\item  Growth by deposition on surfaces\cite{THH}, Edwards/Wilkinson:
\begin{eqnarray}
 \frac{dA}{dt}=  D \cdot \Delta A +  \eta(x,t)
\end{eqnarray}
$A$ stands for the distance to the interface.
$D$ is the surface diffusion constant of the deposited
material and $\eta_i(t)$ is an addititive
noise. 
Other models were proposed by Karkar/Parisi/Zhang, Derrida/Spohn\cite{THH}, etc.
\item Generalised Volterra-Lotka from econophysics:
(Bouchaud\cite{bouch}, Cont, Mezard, Sornette, Solomon\cite{solGLV} etc.)
 \begin{eqnarray}
\frac{dA_i}{dt}=  A_i \cdot \eta_i(t) + \sum_{j} J_{ij} A_j - \sum_{j} J_{ji} A_i
\end{eqnarray}
$A$ stands for individual wealth of agents and $\eta_i(t)$ is a multiplicative
noise.
Agents are involved in binary transactions of "intensity" $J_{ij}$.
 Mean field formal solutions displays scale free distribution of wealth.
Simulations display
patterns on lattice structures (Souma etal\cite{souma}).
\item
Solomon etal\cite{solAB}. Reaction-Diffusion AB models.
\begin{eqnarray}
  \frac{dA}{dt}= k \cdot A \cdot \eta(x,t) + D \cdot \Delta A
\end{eqnarray}
$A$ is the chemical concentration of a
product involved in an auto-catalytic chemical reaction,
$D$ is its diffusion constant. Simulations and formal derivations
yield spatio-temporal patterns similar to
ours. 
 \end{itemize}

\subsection{Variants of the original model}

 We started checking three 
variants, with for instance more realistic production costs
taking into account:
\begin{itemize}
\item 
Influence of capital inertia: production costs don't instantly 
readjust to orders; capital and labour have some inertia
which we modeled by writing that productions costs are a maximum
function of actual costs and costs at the previous period.
\item  
Influence of the cost of credit: production failures increase
credit rates.
\end{itemize}
 The preliminary simulations confirm the genericity of our 
results.

 The third variant is a model with "adaptive firms".
The lattice connection structure
supposes a passive reactive behaviour
of firms. But if a firm is consistently delivering less than
the orders it receives, its customers should order less from it
and look for alternative suppliers.
Such adaptive behaviour leading to an evolutive
connection structure would be more realistic.

 We then also checked an adaptive version of the model
by writing that orders of firm $i$ are proportional 
to the production capacity $A$ of the upstream firms
connected to firm $i$. Simulations
gave qualitative results similar to those obtained 
with fixed structures.

\begin{figure}[htbp]
\centerline{\includegraphics[width=18cm, clip=true, trim= 0 0 0 3]{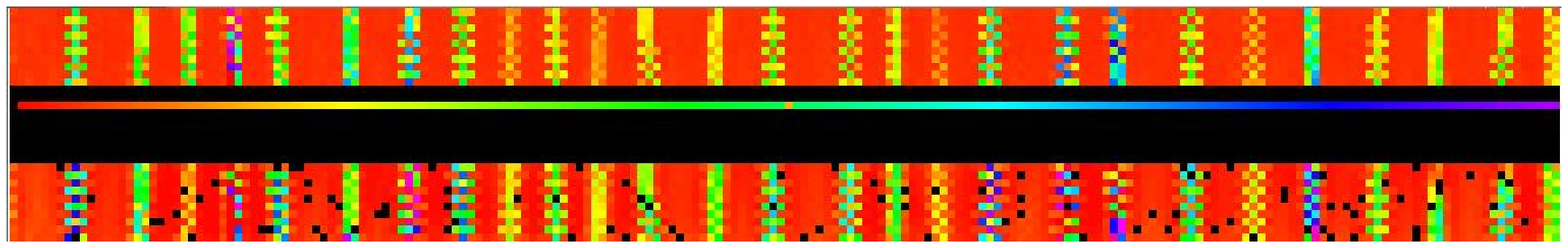}}
\vskip 1cm
\centerline{\includegraphics[width=18cm, clip=true, trim= 0 0 0
  3]{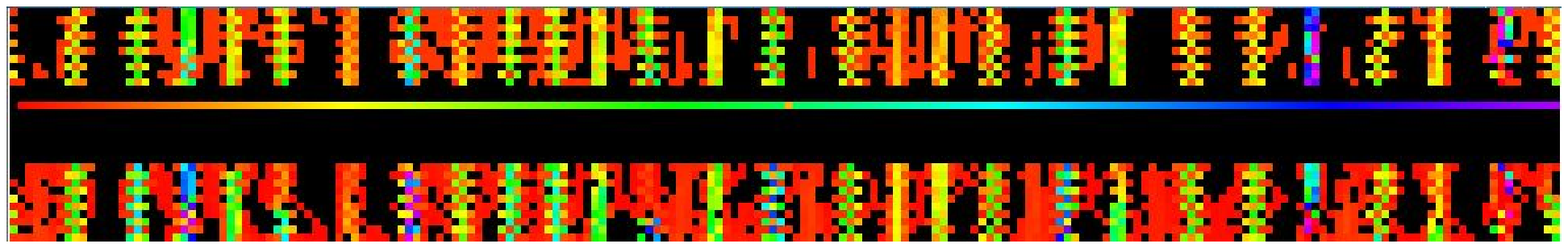}}
\caption {Wealth and production patterns for a network
of "adaptive" firms. The conventions and parameters are the same as for figures
3, 4 and 5, for a 200x10 lattice. Time lag is 1, the two upper patterns correspond to $t=1500$,
the lower ones were taken when $t=1998$.}
\end{figure}

  We observe that adaptation strongly re-enforce the local structure 
of the economy. The general picture is the same
scale free distribution of production and wealth 
with metastable patterns. Due to the strong local character
of the economy:
\begin{itemize}
\item Avalanches of production are observed (see figure 9),
even when time lag is short (time lag of 1).
\item The spatial periodicity of the active zones is increased
(see figure 9 with larger density of smaller zones). But again the activity 
distribution among zones
is like "winner takes all" (figure 7).
\end{itemize}

\subsection{Checking stylised facts}

 Even though the present model is quite primitive\footnote{
We e.g. discuss a "Mickey Mouse" economy with fixed prices
independent from supply and demand. Introducing
price dynamics is not a major challenge: we would simply face
an "embarras de richesse" having to choose among either local or
global prices. In fact both kind of adjustment
have already been tested: global adjustment 
in the case of production cost connected to
production failure through credit costs,
or local adjustment in the case of adaptive behaviour.
We have already shown that they don't change the generic
properties of the dynamics.}
it is still tempting to draw some conclusions
that could apply to real economies.
The most striking result to our mind is the strong 
and relatively stable spatial disparities
that it yields. Let us compare
this prediction to the economic situation of
developing countries: large and persistent disparities in 
wealth and production level as compared to developed countries.
We can even go further and raise questions about the influence
of the depth of the production network or the kind of 
investment needed:
\begin{itemize}
\item One generally agrees that disparities between
developing and developed countries increased since industrial
revolution. This is also a period during which production
became more specialised, which translates in our model 
as increasing the network depth: for instance a shoemaker 
would in the past make and sell a pair of shoes from
leather  obtained from a cattle breeder. Nowadays the shoe production
and delivery  process involve many more stages. Our simulations
have shown that increasing depth increases the fragility 
of economies to failures and bankruptcies. The new industrial
organisation may have detrimental effects on developing economies.
\item 
  Obviously investment policies in developing countries yield 
some coordination across the whole production chain.
Bad economic results might be due to very local conditions but
can also reflect the lack of suppliers/producers connections. 
\end{itemize}

  The above remarks are not counter-intuitive and these conclusions
could have been reached by verbal analysis. What is brought
by the model is the dramatic and persistent consequences of
 such apparently trivial details.

Acknowledgments: We thank Bernard Derrida and Sorin Solomon for
illuminating discussions and the participants to CHIEF Ancona
Thematic Institute, especially Mauro Gallegati. CHIEF
was supported by EXYSTENCE network of excellence,
EC grant FET IST-2001-32802. This research was also supported by
COSIN FET IST-2001-33555, E2C2 NEST 012975 and CO3 NEST 012410
 EC grants.

\end{document}